Topological skyrmions in monolayer multiferroic MoPtGe$_2$S$_6$

Zuxin Fu[1], Kuanrong Hao[4], Min Guo[2], Jingjing He[3], Xiaohong Yan[2], Yangbo Zhou[1*], Lei Shen[5*], and Jiaren Yuan[1*]

[1] School of Physics and Materials Science, Nanchang University, Nanchang 330031, China

[2] School of Physics and Electronic Engineering and School of Materials Science & Engineering, Jiangsu University, Zhenjiang 212013, China

[3] College of Information Science and Technology, Nanjing Forestry University, Nanjing 210037, China

[4] National Center for Nanoscience and Technology, No.11 ZhongGuanCun BeiYiTiao, Beijing 100190, China

[5] Department of Mechanical Engineering, National University of Singapore, 9 Engineering Drive 1, Singapore 117575, Singapore

Corresponding authors. E-mail: yangbozhou@ncu.edu.cn (Y.Z.), shenlei@nus.edu.sg (L.S.), jryuan@ncu.edu.cn (J. Y.)

Two-dimensional (2D) multiferroic materials with coexisting ferroelectricity and ferromagnetism have garnered substantial attention for their intriguing physical properties and diverse promising applications in spintronics. For example, multiferroic materials with electronically controlled broken central symmetry provide a versatile platform for designing and manipulating topological skyrmions and diverse spintronic applications. Here, we investigate the complex magnetic properties of room-temerature multiferroic material MoPtGe$_2$S$_6$ and its electrical control of topological skyrmions using first-principles calculations and atomistic micromagnetic simulations. A sizable Dzyaloshinskii-Moriya interaction (DMI) (2.1 meV) is found in the multiferroic material MoPtGe$_2$S$_6$ with an electrically polarized ground state. The magnetic skyrmions can be stabilized in monolayer MoPtGe$_2$S$_6$ under zero magnetic field, and the chirality of skyrmions can be reversed with electric field-induced flipping of electrical polarization due to the reversed chirality of the DMI. Furthermore, an external magnetic fielc can reverse the magnetization direction and topological charge of the skyrmions as well as tune the size of skyrmions. These results demonstrate that the monolayer MoPtGe$_2$S$_6$ can enrich the 2D skyrmion community and pave the way for electronically controlled spintronic devices.

I. INTRODUCTION

Topologically protected skyrmions are promising as basic memory units due to its nanometer size and small driving current [1]. Skyrmion-based storage, logic, brain-like, and other information devices have been proposed, providing broad application prospects for the development of new memory with high density and low power consumption [2, 3]. Sizable DMI is a crucial factor in generation, stabilization, and manipulation of topological skyrmions [4, 5], which mainly exist in magnetic bulk and thin film materials with symmetry-breaking.[6, 7] Recently, a variety of two-dimensional ferromagnetic materials have been experimentally discovered, such as transition metal halides [8, 9], MXenes [10 ,11, 12], $MnSe_2$ [13, 14, 15], $VSe_2$ [16], $Cr_2Ge_2Te_6$ [17, 18] and $Fe_3GeTe_2$ [19, 20, 21], which provides an ideal platform for spintronics applications. However, most of the two-dimensional ferromagnetic materials reported in the experiment have the absence of inversion symmetry breaking, which is a crucial factor for DMI [22] and skyrmion creation. Several strategies to break the structural inversion symmetry have been proposed to address this limitation, including constructing Janus monolayer structures [23, 24, 25, 26], applying perpendicular electric fields [27] and stacking vdW heterostructures. These schemes can effectively tune the electric potential, providing a new path to induce DMI and mediate the magnetic interactions and skyrmions. Moreover, ferroelectric polarization is another flexible solution to induce DMI [28].

Recently, two-dimensional ferroelectric materials have attracted wide attention for the potential applications in non-volatile nano-devices.[29, 30, 31] Two-dimensional van der Waals ferroelectric materials include 1T phase transition metal chalcogenides, indium triselenide ($In_2Se_3$), copper indium phosphorus sulfur ($CuInP_2S_6$) [32, 33, 34]. Furthermore, certain two-dimensional materials exhibit concurrent intrinsic ferromagnetic and ferroelectric properties, named multiferroics, such as $CuCrP_2S_6$ and $NiI_2$.[35, 36] In such multiferroic materials, a strong coupling exists between charge, spin, orbital, and lattice degrees of freedom.[37] The external electric field will induce orbital reconstruction and charge transfer, which enables electric field control of magnetic properties and provides the possibility for the design of new multifunctional magnetoelectric memory devices [38]. The noncentrosymmetric nature allows DMI and skyrmions to hold in multiferroic materials.[39, 40, 41] However, two-dimensional multiferroic materials with intrinsic high Curie temperature and out-of-plane ferroelectric polarization are rare. [42] Some external methods/techniques are required to realized them, such as doping elements [] or adding substrates [].

In this work, we investigated the magnetic properties of monolayer $MoPtGe_2S_6$ by first-principles calculations and atomic spin model simulations. It is found that monolayer $MoPtGe_2S_6$ is a multiferroic semiconductor with coexisting ferroelectricity and ferromagnetism. The ferroelectricity is from the Ge-Ge atomic displacement, and the ferromagnetism originates from the super exchange interaction between neighboring Mo atoms. The spontaneous shift of Ge-Ge pairs induces spontaneous electric polarization (0.09 e Å) along with geometric symmetry breaking, where significant DMI can be induced, which is enough to stabilize the spontaneous skyrmions without magnetic fields. Importantly, the chirality of DMI and skyrmions can be reversed via ferroelectricity polarization induced by an external electric field. Moreover, the manipulation of topological charge and size of skyrmions can be achieved through an external magnetic field. Monolayer MoPtGe$_2$S$_6$ thus provides an excellent platform for realizing tunable skyrmion-based spintronic devices.

## II. COMPUTATIONAL METHODS

We performed first-principles calculations on MoPtGe$_2$S$_6$ monolayer properties using the Vienna *ab initio* simulation package (VASP)[43] based on the density-functional theory (DFT) framework. The projector augmented wave (PAW) method [44] was employed to describe the interaction between the cores and valence electrons. The generalized gradient approximation (GGA) of Perdew-Burke-Ernzerhof (PBE) was chosen to treat the exchange-correlation [45]. The MoPtGe$_2$S$_6$ monolayer was constructed with a slab model with a 15 Å vacuum layer. On-site Hubbard U was equal to 2.0 eV and 0.5 eV for Mo-4d and Pt-5d electrons, as reported in a previous study[46]. The cutoff energy for the plane wave was set as 500 eV. In order to get accurate magnetic parameters, the energy and force convergence criteria were set as $10^{-6}$ eV and 0.001eV/Å.

Atomistic micromagnetic simulations were performed using the software package VAMPIRE [47], and the magnetic parameters of the model were obtained from DFT calculations for MoPtGe$_2$S$_6$. We solved the Hamiltonian of the magnetic system using the Landau-Lifshitz-Gilbert equation. The LLG was given by

$$\frac{\partial \mathbf{S}_i}{\partial t} = -\frac{\gamma}{(1+\lambda^2)}\left[\mathbf{S}_i \times \mathbf{B}^i_{\text{eff}} + \lambda \mathbf{S}_i \times \left(\mathbf{S}_i \times \mathbf{B}^i_{\text{eff}}\right)\right]$$

where $\mathbf{S}_i$ is a unit vector representing the direction of the magnetic spin moment of the Mo atom, $\gamma$ is the gyromagnetic ratio, $\lambda = 0.1$ is the damping parameter, and $\mathbf{B}^i_{\text{eff}}$ is the net magnetic field on each spin. The system size was set to 120 nm × 200 nm with periodic boundary conditions in the x and y directions. 3,000,000 integration steps and an integration timestep of $10^{-16}$ were adopted to guarantee the accuracy of the results.

III. RESULTS AND DISCUSSION

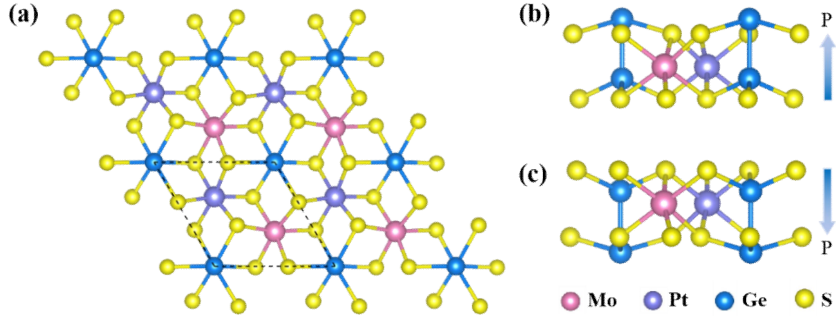

**Fig. 1.** (a) Top view of the MoPtGe$_2$S$_6$ monolayer. (b) Side view of the MoPtGe$_2$S$_6$ monolayer in the ferroelectric polarization up +P and ferroelectric polarization down -P cases.

Figure1 describes the fully optimized structure of monolayer MoPtGe$_2$S$_6$. Monolayer MoPtGe$_2$S$_6$ consists of an S framework bridge with Mo, Pt, and Ge–Ge in a triangular pattern. Mo and Pt atoms are bonded with six S atoms, and each Ge atom is surrounded by three S atoms. Figures1(b) and 1(c) show the FE phase with the +P and -P polarization directions. For the PE phase, Ge-Ge atomic pairs are located between the upper and lower S atoms planes. For the FE phase, Ge-Ge atomic pairs shift along the z-axis, accompanied by spontaneous out-of-plane polarization. When the system transitions from the PE to FE phases, the geometric symmetry will break as the Ge−Ge atomic pair shifts along the z-axis, which also leads to the separation of positive and negative charge centers, resulting in a nontrivial electric dipole as shown in the electrostatic potential distribution diagram [Fig. 2(a)]. Both the relaxed lattice constants and polarization phases are in good agreement with the previous report. The evaluated electric dipole $\varepsilon$ of the monolayer MoPtGe$_2$S$_6$ is 0.09 formula unit (f.u.). The FE polarization is different from 2D FE CuInP$_2$S$_6$ and CuCrP$_2$S$_6$, originating from the shift of metal ions along the normal plane. The formation energy of the monolayer MoPtGe$_2$S$_6$ is calculated to be -0.8 eV per f.u., indicating that it possesses excellent energy stability. The phonon calculations and molecular dynamic simulation of the FE phase are performed to confirm their dynamical stability and kinetic stability. No imaginary frequency is found throughout the entire Brillouin zone (BZ), as shown in the phonon band structure [Fig. 2(b)], which verifies the dynamic stability of monolayer MoPtGe$_2$S$_6$. Furthermore, the FE phase maintains the integrity of the original structures with small fluctuations of energy [Fig. 2(c)], indicating that the structure is thermally stable at 300 K.

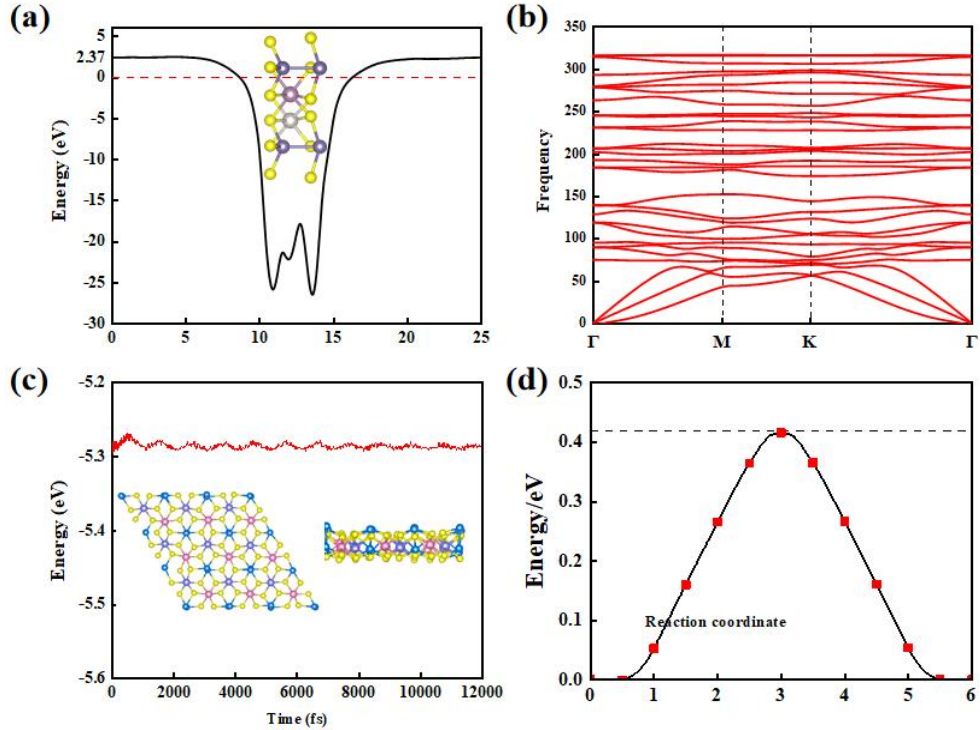

**Fig. 2** (a) The electrostatic potential distribution of the MoPtGe$_2$S$_6$ along the Z-direction, (b) the phonon spectrum of the MoPtGe$_2$S$_6$ monolayer. (c) and (d) molecular dynamics simulation and NEB results.

To understand the transition between polarization up and polarization down, we study the energy path with a climbing-image nudged elastic band (CI-NEB) method, as depicted in Fig. 2(d). The minimum energy barrier from the polarization up to polarization down state is about 0.42 eV/f.u. for monolayer MoPtGe$_2$S$_6$, which is appropriate for facile switching of ferroelectricity between upward and downward polarization via the flipping of external electric field. The electric properties of monolayer MoPtGe$_2$S$_6$ in FE states are investigated. The spin-dependent band structure without SOC for FE phases is depicted in Fig. 3(a). Spin-up and spin-down bands are split with the valence band maximum (VBM) between Γ and M points and the conduction band minimum (CBM) between the Γ and K points. While the CBM originates from the spin-up band, and the VBM stems from the spin-down band. Both the spin-up band with the gap (0.68 eV) and spin-down bands with gaps (0.83 eV) do not pass through the Fermi level, demonstrating that monolayer MoPtGe$_2$S$_6$ is a magnetic semiconductor with an indirect band gap of 0.54 eV, which is consistent with the previous result [hao]. The partial density of states (PDOS) in Fig. 3(b) reveals that Mo and six S atoms contribute predominantly to valence and conduction bands. When SOC is taken into consideration, there is no obvious changes in electronic structures near the Fermi level.

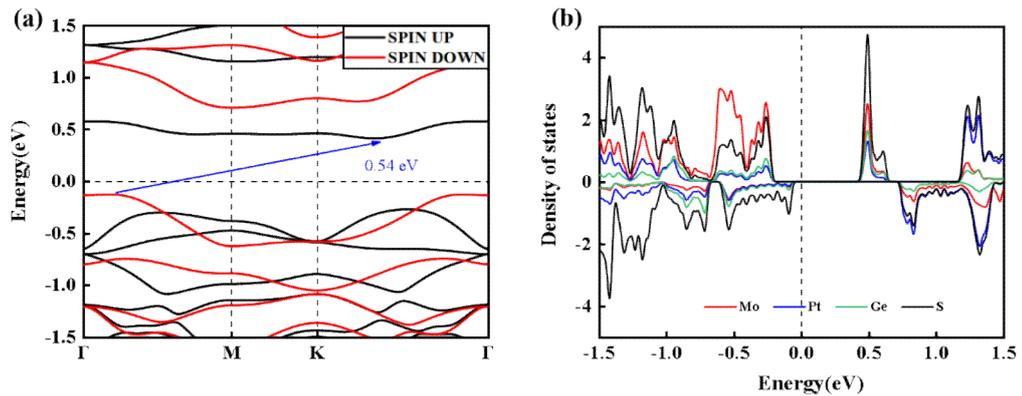

**Fig. 3** The (a) band structure and (b) PDOS of the MoPtGe$_2$S$_6$ monolayer.

The spin density of monolayer MoPtGe$_2$S$_6$ for FM states is shown in Fig. 4(a). One can find that the spin density is mainly distributed on Mo atoms, revealing that the magnetic moment mainly stems from Mo atoms. The Mo atom, with a 2.1 μB magnetic moment, forms a typical triangular magnetic lattice. The effective Hamiltonian of magnetic interactions is comprised of the following three energy terms:

$$H_{spin} = -\sum_{ij} J_{ij} S_i \cdot S_j - \sum_{ij} K_{ij} S_{iz} S_{jz} - \sum_{ij} D_{ij} \cdot (S_i \times S_j)$$

Here, S$i$ (S$j$) refers to the spin vector of the Mo atom at site $i$ ($j$). $J_{ij}$, $K_{ij}$ and $D_{ij}$ stand for the parameters of Heisenberg isotropic exchange coupling, single-ion magnetic anisotropy, and DMI. Note that the SOC interaction should be considered in the calculations of the last two items.

The collinear magnetic order was investigated using a 2 × 2 supercell, considering both ferromagnetic (FM) and antiferromagnetic (AFM) states in the total energy calculation. The evaluated exchange interaction parameter, as listed in Table 1, is up to 35.12 meV, indicating that FM exchange interaction is favorable between the neighboring Mo atoms. The strong FM ordering stems from the indirect exchange interaction between the nearest neighboring magnetic Mo atoms in the triangular magnetic lattice, mediated by the nonmagnetic Pt and S atoms. The connecting bridges between the neighboring magnetic Mo atoms consist of nonmagnetic S−Pt−S units, with a substantial distance of 6.31 Å, indicating the presence of "super-super-exchange interactions." Hence, the coexistence of FM and ferroelectric (FE) phases in monolayer MoPtGe$_2$S$_6$ is confirmed. The magnetic anisotropy energy is evaluated as the energy difference between the in-plane (100) and out-of-plane (001) magnetization directions. The calculated magnetic anisotropy energy is -0.47 meV, indicating the monolayer MoPtGe$_2$S$_6$ tends to align along out-of-plane.

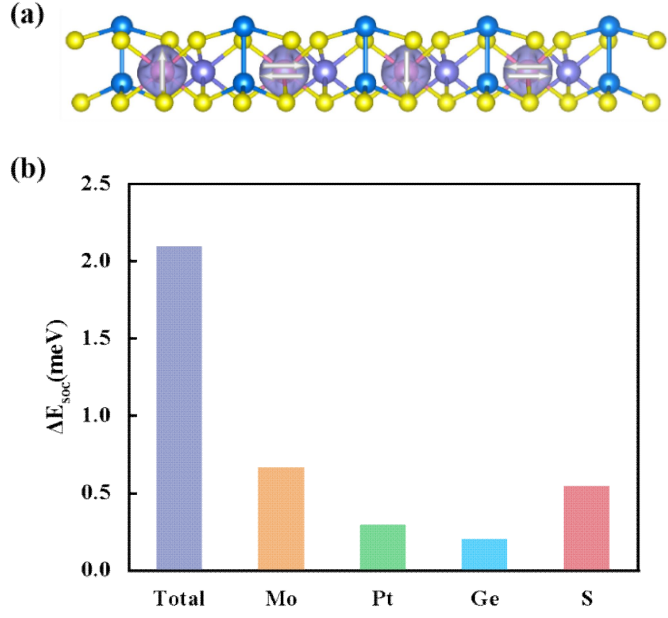

**Fig 4.** (a) The spin density and left-handed and right-handed spin configurations of monolayer MoPtGe$_2$S$_6$. (b) The atomically resolved localization of the associated SOC energy, $\Delta E_{soc}$.

Dzyaloshinskii-Moriya interaction (DMI) governs the formation of magnetic skyrmions. According to Moriya symmetry rules, the vector of $D_{ij}$ can be expressed as $D_{ij} = d_{\parallel}(z \times u_{ij}) + d_{\perp}z$, where $z$ and $u_{ij}$ denote the unit vector along the $z$ direction and from the site $i$ to site $j$, respectively. It is stressed that the out-of-plane component $d_{\perp}$ has a negligible impact on the formation of skyrmion. As a result, only the in-plane component $d_{\parallel}$ was taken into account in the subsequent computation. Two noncollinear magnetic orders with left-handed and right-handed spin spirals are considered in a 4×1×1 supercell, as shown in **Fig. 4(a)**. Interestingly, the strength of the in-plane component $d_{\parallel}$ in the monolayer MoPtGe$_2$S$_6$ is up to 2.1 meV (right-handed spin spirals) for polarization up, which could be comparable to that in previously reported state-of-art ferromagnetic/heavy metal multilayers []. The relevant atomic resolved SOC energy difference ΔEsoc is depicted in **Fig. 4(b)** to clarify the origin of DMI in the monolayer MoPtGe$_2$S$_6$. It is evident that the predominant contribution to DMI originates from the magnetic Mo and nonmagnetic S atoms. The underlying mechanism for such a large DMI value in MoPtGe$_2$S$_6$ is related to the inversion symmetry broken induced by ferroelectric polarization. The external electric field can flip the electrical polarization of the multiferroic system, thus changing the magnetic parameters due to the strong magnetoelectric coupling. As listed in Table 1, the DMI interaction reverses from 2.1 meV (right-handed spin spirals) to -2.1 meV (left-handed spin spirals) along with the flip of electrical polarization induced by an external electric field.

TABLE I. The lattice parameters, total magnetic moment, Heisenberg isotropic exchange coupling, single-ion magnetic anisotropy, and DMI for MoPtGe$_2$S$_6$ monolayer.

|  | a=b (Å) | M$_{tot}$ (μB) | J (meV) | d$_{//}$ (meV) | MAE (meV) |
|---|---|---|---|---|---|
| MoPtGe$_2$S$_6$ (P↑) | 6.31 | 2.28 | 35.12 | 2.1 | -0.47 |
| MoPtGe$_2$S$_6$ (P↓) | 6.31 | 2.28 | 35.11 | -2.1 | -0.47 |

Employing our DFT calculated magnetic parameters J, K, and $d_\parallel$, we perform atomistic spin simulations of monolayer MoPtGe$_2$S$_6$ using VAMPIRE [51]. The magnetization is initialized to a random state with the periodic boundary, and the relaxed states are shown in Fig. 5(a) with $d_\parallel = 2.1\text{eV}$ and 5(b) with $d_\parallel = -2.1\text{eV}$ without magnetic field. One can find that monolayer MoPtGe$_2$S$_6$ exhibits stable Neel skyrmions, in which the magnetization at the centre is along the z direction, opposing the ferromagnetic background at the boundary with magnetization along the -z direction in both cases. However, the spiral directions are different for $d_\parallel = 2.1\text{eV}$ and $d_\parallel = -2.1\text{eV}$. When the electrical polarization of monolayer MoPtGe$_2$S$_6$ flips from polarization up to polarization down, the chirality of the skyrmions changes from right-handed [Fig. 5(a))] to left-handed spin spirals [Fig. 5(b)] due to the reversal in DMI chirality. The magnetic skyrmions exhibit topological stability, characterized by a quantized topological charge $n$ which is defined as $n = \frac{1}{4\pi}\int \left(\frac{\partial m}{\partial x} \times \frac{\partial m}{\partial y}\right) \cdot m \, \text{dxdy}$, where $m$ is the normalized magnetization vector. The topological charge of skyrmions of MoPtGe$_2$S$_6$ is equal to one under a zero magnetic field. Hence, polarization flipping changes neither $m$ nor the sign of topological charge for the skyrmion as shown in **Figs. 5(a)** and **Fig. 5(b)**. The diameter of the skyrmion, *R*, is easily determined by reading the out-of-plane magnetization component $m_z(x)$. The diameter of skyrmion at the zero magnetic field for monolayer MoPtGe$_2$S$_6$ is approximately 9.46 nm.

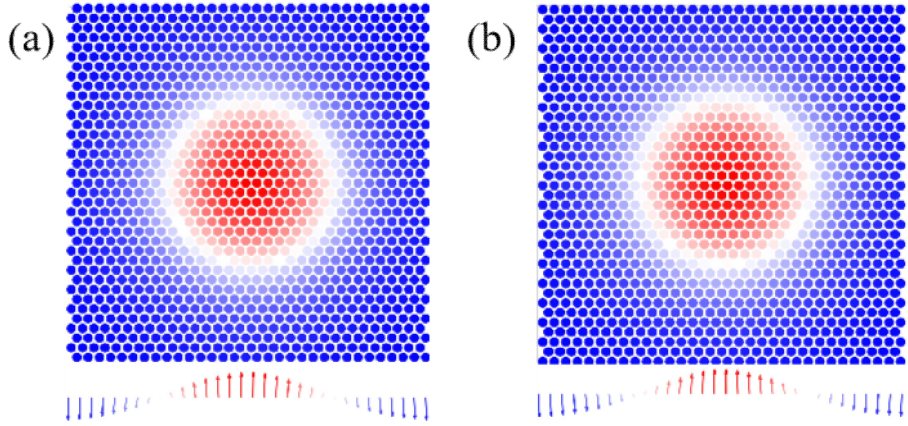

**Fig 5**. Magnetization distribution for the relaxed states at zero magnetic field in MoPtGe$_2$S$_6$  (a) skyrmion profile of polarization up state +P in the xy plane and xz plane, respectively. (b) skyrmion profile of polarization up state -P in the xy plane and xz plane, respectively.

Next, we proceed with the external magnetic field effect on the skyrmions in monolayer MoPtGe$_2$S$_6$. The magnetic states under negative external magnetic fields are depicted in **Fig. 6** and **Fig. 7**. The magnetic state exhibits minimal changes under small negative magnetic fields for both polarization up and polarization down in monolayer MoPtGe$_2$S$_6$. Take the ferroelectric polarization up state as an example, one can see that the skyrmion rim rapidly shrinks to its center and the size of skyrmions continues to decrease with the increasing negative magnetic field. The size of skyrmions diminishes to 5.68 nm at -0.8 T and eventually completely becomes a ferromagnetic state with all spins aligned in the negative direction at -0.9 T. The nontrivial skyrmion state transitions to a trivial ferromagnetic state with the topological charge changing from n = 1 to n = 0.

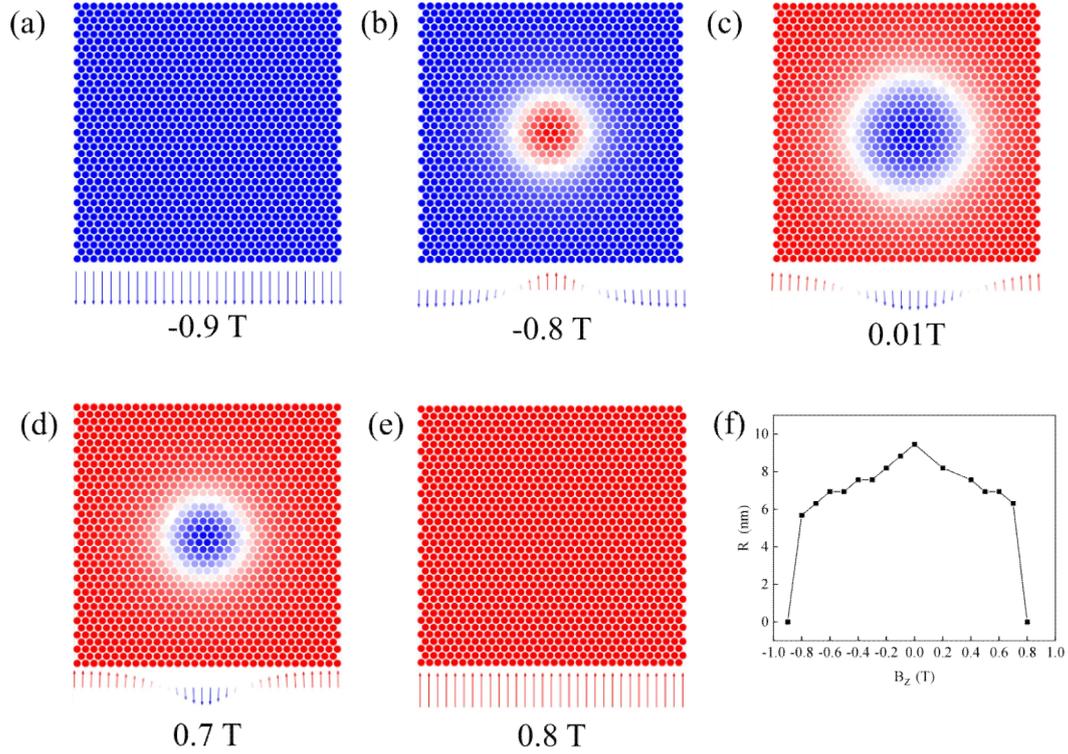

**Fig. 6**. Magnetization distribution for polarization up state +P of MoPtGe$_2$S$_6$ at different magnetic field. (a) Magnetization distribution at -0.9 T magnetic field. (b) Magnetization distribution at -0.8 T magnetic field. (c) Magnetization distribution at 0.01 T magnetic field. (d) Magnetization distribution at 0.7T magnetic field. (e) Magnetization distribution at 0.8 T magnetic field. (f)The size of the skyrmions as function as magnetic field.

When a positive magnetic field is applied, the magnetization direction of the skyrmion core and the surrounding ferromagnetic background after full relaxation, as shown in Fig. 6 and Fig. 7, are reversed for both polarization up and polarization down in monolayer MoPtGe$_2$S$_6$, which changes the sign of topological charge *n* from 1 to -1, even at a very small positive magnetic field (0.01T). However, the chirality of skyrmions is still right-handed and left-handed spin spiral for polarization up and polarization down, which has not changed compared to the chirality of magnetic states at zero magnetic field and negative magnetic field. Similar to the scenario under a negative magnetic field, the skyrmion size decreases with the increase of external positive field, reaching 6.31 nm at 0.7 T. Upon reaching a magnetic field of 0.8 T, the external field disrupts the state of skyrmions, ultimately transforming it into a FM state with all spins aligned in the positive direction.

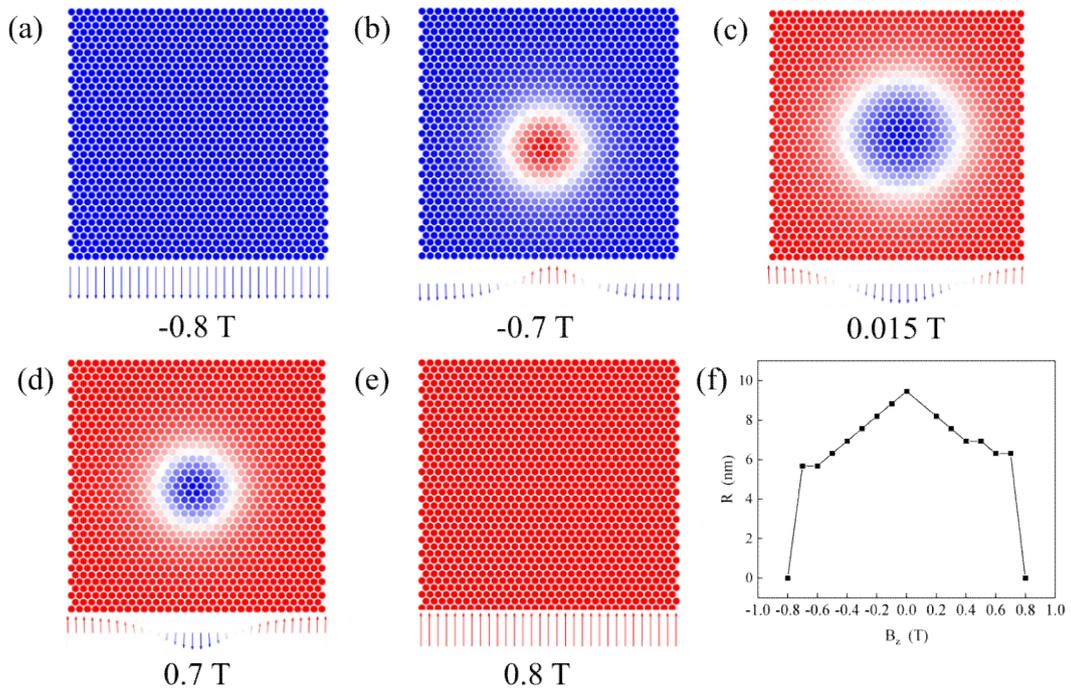

**Fig 7**. Magnetization distribution for polarization down state -P of MoPtGe$_2$S$_6$ at different magnetic field. (a) Magnetization distribution at -0.8 T magnetic field. (b) Magnetization distribution at -0.7 T magnetic field. (c) Magnetization distribution at 0.01 T magnetic field. (d) Magnetization distribution at 0.7T magnetic field. (e) Magnetization distribution at 0.8 T magnetic field. (f)The size of the skyrmions as function as magnetic field.

In multiferroic material MoPtGe$_2$S$_6$ monolayer the electric field first flips the electric polarization along with DMI sign which then reverse the chirality of the skyrmion. In order to change the sign of topological charge for skyrmions, an external magnetic field is needed to introduce external forces which disrupt the delicate balance of all associated energies and reverses the magnetization direction of the skyrmion core. Simultaneously, the DMI competes with the Zeeman energy to stabilize skyrmion and prevent their collapse into the FM state.

## IV. CONCLUSION

In summary, we study the non-collinear magnetic properties of multiferroic MoPtGe$_2$S$_6$ monolayer, by combining first-principles calculations and atomistic spin model simulation. Our results indicate that a sizable DMI is found in multiferroic monolayer MoPtGe$_2$S$_6$, which stabilizes the topologically nontrivial magnetic skyrmions with nanometer size under zero magnetic field. The electric-field switch of the chirality of skyrmions and DMI is realized in MoPtGe$_2$S$_6$. The magnetization direction and topological charge of the skyrmion can be reversed under a positive magnetic field, and the size of skyrmions can be tuned by an external magnetic field. The multiferroic monolayer MoPtGe$_2$S$_6$ holds promise for designing and developing an electric-field and magnetic-field control of topological magnetism.

**ACKNOWLEDGMENT**

This work was supported by the National Key Research and Development Program of China (Grant No. 2022YFA1405200) the National Natural Science Foundation of China (Grant No. 12174158, 62264010, 12264026, 12004142, 62201268), the Natural Science Foundation of Jiangxi Province, China (Grant No. 20212BAB211023, 20224BAB211013), the Innovation and Entrepreneurship Leading Talent Plan of Jiangxi Province (Grant No. Jxsq2023101068) and Singapore Ministry of Education Academic Research Fund Tier 1 (Grant No. A-8001194-00-00).